# A study of small impact parameter ion channeling effects in thin crystals

M. Motapothula[1,2], M. B. H. Breese[1,3+]

[1] Center for Ion Beam Applications, Physics Department, National University of Singapore,
Singapore 117542
[2] NUSNNI-NanoCore, National University of Singapore,
Singapore 117576
[3] Singapore Synchrotron Light Source (SSLS), National University of Singapore,
Singapore 117603

## Abstract

We have recorded channeling patterns produced by 1 to 2 MeV protons aligned with <111> axes in 55 nm thick silicon crystals which exhibit characteristic angular structure for deflection angles up to and beyond the axial critical angle, $\psi_a$. Such large angular deflections are produced by ions incident on atomic strings with small impact parameters, resulting in trajectories which pass through several radial rings of atomic strings before exiting the thin crystal. Each ring may focus, steer or scatter the channeled ions in the transverse direction and the resulting characteristic angular structure beyond $0.6\psi_a$ at different depths can be related to peaks and troughs in the nuclear encounter probability. Such 'radial focusing' underlies other axial channeling phenomena in thin crystals including planar channeling of small impact parameter trajectories, peaks around the azimuthal distribution at small tilts and large shoulders in the nuclear encounter probability at tilts beyond $\psi_a$.

[+] Contact author: phymbhb@nus.edu.sg

# 1. Introduction

*1.1. Channeling in thin crystals*

Channeling in crystals occurs when the transverse energy of incident ions is less than the maximum potential energy associated with an atom row or plane [1-4]. Lindhard used a continuum approximation [1] to derive a critical angle, $\psi_a$, for axial channeling for ions incident at small angles to major axes. Axially aligned ions which enter close to the minimum of the potential well typically have trajectories confined within the same channel throughout their path, whereas ions incident closer to atomic strings may enter adjacent potential wells and wander between channels [5,6]. Ions incident at tilts slightly greater than $\psi_a$ travel close to the atomic rows and so are strongly scattered, i.e. they have blocking trajectories [7,8], with an increased backscattered ion or nuclear reaction yield above that measured in a non-channeled alignment.

Channeling analysis in conjunction with RBS (Rutherford Backscattering Spectrometry) is used to provide quantitative information on crystallinity close to the surface by measuring the minimum yield, defined as the ratio of backscattered counts at channeled and random alignments from depths just below the surface [2-4]. With sufficiently-high depth resolution, multiple surface peaks are observed in RBS-channeling spectra. Barrett [9] developed a Monte Carlo computer code which modelled channeling phenomena and derived the contribution to the nuclear encounter probability (NEP) versus depth [10]. He demonstrated that NEP peaks occur at depths corresponding to where axially aligned ions are radially focused onto atomic strings in concentric rings around a central string on to which they impact. Bøgh (chapter 15 of Ref. [2]) presented high resolution RBS-channeling measurements of scattered proton and helium ion yields versus depth in Si where the observation of a resolved sub-surface peak was consistent with Barrett's predictions based on the radial focusing effect.

Fig. 1 shows the NEP versus depth for 2 MeV protons aligned with the Si <111> axes, calculated using the FLUX code which utilizes the Ziegler-Biersack-Littmark universal potential and a binary collision model with an impact parameter dependent algorithm for energy loss [11,12]. The first three NEP peaks are around depths of 40, 95 and 160 nm. The contributions to the total NEP for beam components with impact parameters of $b < 0.2$ Å and for $b = 0.2$ to 0.4 Å are also indicated, scaled to their respective fraction of the geometric area within the unit cell. While only 2.4% of incident ions have $b < 0.2$ Å this component dominates the NEP for the full beam distribution. Smaller NEP peaks are observed to a depth of 500 nm, with an amplitude decreasing with depth in a manner similar to the NEP oscillations produced at (110) planar alignment, also shown in Fig. 1. This is further discussed in section 3.3.

'Star patterns' produced by MeV ions passing through crystals, typically tens of microns thick, were utilized early in the investigation of ion channeling phenomena [13]. In thick crystals the beam is in statistical equilibrium and multiple scattering dominates the exit angular distribution. When aligned with a major axis the central angular region is uniformly filled, as are planar directions radiating away from the axis; at larger deflection angles these planar directions appear unfilled due to blocking effects.

At the other extreme of crystal thickness studies of rainbow channeling and superfocusing effects use crystals only a few hundred nanometers thick. The hyperchanneled beam fraction, produced by ions incident on the central region of each potential well where $b > 0.7$ Å, has deflection angles of less than $\sim 0.25\psi_a$ [14-18].

Armstrong et al [19] presented channeling patterns produced by 4 MeV protons transmitted through Si crystals between 0.5 to 1.1 µm thick at small tilts to <111> and <100> axes, where ring-shaped "doughnut" patterns were recorded. Rosner et al [20] recorded doughnut patterns produced by 3.2 MeV protons transmitted through Si crystals of 190 nm thick near <110> axes. The angular radius of the doughnut corresponds to the angle between the beam direction and the crystal axis and the doughnut has a continuous azimuthal intensity modulated by the symmetry of the axis such that planar directions intersecting the axis appeared dark.

Only recently have ultra-thin Si membranes thinner than 100 nm become available for ion channeling analysis [21,22]. Channeling patterns from such membranes exhibit minimal blurring due to multiple scattering which contributes a calculated angular spread of only ~0.01° for a 2 MeV proton beam transmitted through a 55 nm Si layer, compared to typical values of $\psi_a = 0.3°$ to $0.4°$. This allows observation of faint and fine angular structure [23,24] and different channeling behaviour at wide and narrow {111} planes [25].

*1.2. Summary of this work*
We use Barrett's work [10] as a starting point in our study relating radial focusing effects of MeV protons transmitted through ultra-thin Si crystals to the resulting characteristic angular structure and the NEP depth distribution. We concentrate on the NEP peak at 95 nm for 2 MeV protons along <111> axes being the largest and similar to the path length through our ultra-thin membranes. The term 'radial focusing' is used to encompass all three possible effects of focusing, steering and scattering. We show that radial focusing underlies a range of ion channeling phenomena. First, it is generally considered that axially channeled ions which are scattered towards major planar directions are excluded from travelling parallel to them, i.e. they cannot be channeled. We show that the underlying scattering mechanism involves a well-defined sequence of interactions by successive radial rings which deflect ions towards major planar directions where they undergo repeated scattering from strings along these directions, producing a range of transverse angles. Those ions with large transverse energies are non-channeled or blocked but for many ions with lower transverse energy such motion can be considered as large amplitude planar channeling. The outcome is that many ions are deflected further radially outwards along these directions than in other azimuthal directions.

Second, in axial channeling patterns from crystals thinner than 100 nm the azimuthal angular distribution for tilts of $< \psi_a$ comprises resolved peaks, instead of a continuous rings as observed in thicker crystals [19,20]. We show that these are produced by the radial focusing effect acting on a large beam fraction owing the crystal tilt. Such resolved angular peaks are thus produced by the same process as those observed at axial alignment for small $b$ components and exhibit similar depth and tilt behavior.

Third, we show that faint angular structure is present in channeling patterns at deflection angles beyond $\psi_a$. This is produced by the beam component with $b$ comparable to the thermal vibration amplitude of the atomic strings. Such angular structure is too faint to be experimentally observed in channeling patterns at axial alignment but the predicted features are observed at small tilts owing to the larger beam fraction contributing to it.

Fourth, we show that radial focusing produces large peaks in the NEP, up to five times the non-channeled level, for tilts just beyond $\psi_a$ along the same minor planar directions as are responsible for peaks in the NEP for axial alignment. This effect is strongly depth dependent, occurring only over the narrow range corresponding to the NEP peaks at axial alignment. This explains all previous observations of large shoulders measured in the backscattered ion or nuclear reaction yield at such tilts.

## 2. Experimental and Simulation Parameters

*2.1. Experimental Details*
Ultra-thin membranes were prepared as described in [21] and mounted on a goniometer in the chamber of a nuclear microprobe [26]. The MeV proton beam was focused to a spot size of ~1 µm with a beam convergence angle of ~0.01°. Channelling patterns were recorded by photographing a high-sensitivity aluminium-coated YAG (yttrium aluminium garnet) scintillator screen located 70 cm downstream of the crystal, using a beam current of about 1 to 10 pA. The digital camera exposure time was varied from 0.1 secs to 20 secs, depending on whether one is trying to record the intense, central region of the patterns using a short exposure or the fainter, outer regions requiring a longer exposure, resulting in the central region becoming saturated. Different exposures were recorded to capture all angular features.

*2.2. Channeling Parameters*
The <111> axis has wide, open channels, allowing ions with sufficient radial energy to wander freely between adjacent potential wells. Ions which are incident close to an atomic string are sensitive to the discrete nature of the string potential owing to the individual atoms along it and also their thermal vibrations. In Si the two dimensional *r.m.s.* thermal vibration amplitude of the atomic strings at room temperature, $\rho_{th}$ = 0.106 Å [3] and the spacing between <111> strings is 1.9205 Å, so impact parameters of $b < \rho_{th}$ comprise ~0.6% of the incident beam.

Due to the discrete nature of the string potential, one should consider the effective screening radius instead of original Thomas-Fermi ($a_{TF}$) radius, $a' = \left(a_{TF}^{-2} + \rho_{th}^{-2}\right)^{-\frac{1}{2}} = 0.089$ Å [31]. These factors determine the minimum distance of approach for which an atomic string is able to produce the required correlated sequence of scattering for channeling, i.e. for which transverse energy is conserved and the continuum potential model is valid. The characteristic angle for axial channeling [4] is:

$$\psi_a = \frac{2Z_1 Z_2 e^2}{Ed}$$

where $Z_1$, $Z_2$, are the atomic numbers of the incident ions and the lattice atoms, $e^2$ is the electronic charge squared, $E$ is the beam energy and $d$ is the atom spacing along the string. For 2 MeV protons along the Si <111> axis, $\psi_a = 0.41°$. Our measured value of the half-width-at-half-maximum, $\psi_{1/2}$ for 2 MeV protons along the Si <111> axis is 0.36°, i.e. typically 10% lower than $\psi_a$ (page 43, Ref. [4]).

*2.3. <111> Potential Distribution*
Figure 2 shows a map of the static atomic potential at the Si <111> axes for protons averaged along the ion direction. The central string upon which all ions are incident in our simulations is shown as a black dot in the lower left corner and the first twelve radial rings of atomic strings around this central string are indicated. Rings shown in purple correspond to minor planar directions; these are of particular importance as producing strong scattering of radially focused beam, as further discussed below.

*2.4. FLUX simulations and image processing*
FLUX simulations are used to study how different impact parameters *b* are related to features in the transmitted angular distribution and to link such features to the NEP depth distribution. Our simulations use the ZBL potential which gives the best agreement in the observed angular structure with experimental results corresponding to the same depth. However, simulations using the Molière or HF potentials give the same observed changes in angular distribution but with a different depth dependence.

Barrett [10] plotted a small number of individual ion trajectories viewed perpendicular to the surface in order to show the focusing effect of successive radial rings as the ions moved radially outwards away from their impact point on a central string. While this approach enables individual trajectories to be followed as they pass through different rings at different depths, one can only view a limited number of trajectories. We use a different approach here which allows high statistics simulations of many (up to 25 million) ion trajectories to generate maps of the spatial and angular intensity distributions at specific depths. This approach also provides a closer analogy to the experimental observation of a channeling pattern showing the angular distribution after having passed through a specific crystal thickness.

The use of large statistics allows image regions of low spatial and angular intensity in the simulations results to be resolved by progressively saturating images. Low saturation means that the range of intensities plotted varies from 0 to the maximum value normalized to 1. Medium/high/very-high saturation mean that the maximum value plotted is reduced by successive orders of magnitude, above which areas are of uniform colour.

### 3. Results

*3.1. Full beam distribution*

FLUX simulations were performed to generate the angular distribution of 2 MeV protons transmitted through different depths along the Si <111> axis, with beam uniformly distributed over the full unit cell. A full set is given in Fig. S1 and a selection is shown in Fig. 3 for medium, high and very-high saturation. At a depth of 80, 95 nm in Fig. 3a an angular region at ~$0.6\psi_a$ comprising six resolved dots aligned along {112} directions observed. At higher saturation another angular region extending out to ~$0.8\psi_a$ is observed comprising radiating lines. By a depth of 130 nm the angular structure comprises an intense, non-continuous azimuthal ring at ~$0.6\psi_a$ and a faint outer ring at ~$0.8\psi_a$, indicated by an arrow.

At very high saturation, Fig. 3c, there is another angular region (see arrow at depth of 80 nm) at deflection angles up to ~$1.3\psi_a$ where faint, angular structure is observed up to depths of ~100 nm. At greater depths such angular structure disappears. This faint region is further discussed after presenting simulations of the angular distribution for $b < \rho_{th}$ in section 3.4.

Experimental channeling patterns were recorded to see whether the above simulated features could be observed. Fig. 4 shows patterns recorded for 1 and 2 MeV protons transmitted through a 55 nm thick [001] Si membrane aligned with the <111> axis, giving a path length of 95 nm. For 2 MeV protons this corresponds to the depth of the large NEP peak. In the patterns produced by 2 MeV protons in Fig. 4b a region extending up to ~$0.6\psi_a$ comprising six resolved dots aligned along {112} directions is observed and with increasing saturation an outer region extending up to ~$0.8\psi_a$ containing the same characteristic angular structure as in Fig. 3 at this path length.

The NEP depth scale varies as the square root of the beam energy, so the incident beam energy can be varied to change the depth of the NEP peaks. Thus for 1 MeV protons a path length of 95 nm is similar to what is expected for 2 MeV protons at a path length of ~134 nm, corresponding to the trough after the second NEP peak. In Fig. 4a at medium saturation, the same characteristic ring-like angular structure is experimentally observed for 1 MeV protons as in simulations for 2 MeV protons at a depth of 130 nm.

We have demonstrated good agreement between the simulated and experimental angular distributions for the full beam distribution. Different angular regions are observed depending on the image saturation. We next consider which $b$ components are associated with each angular region in the patterns.

*3.2. Medium impact parameters of 0.2 Å < b < 0.6 Å*

Fig. 5 shows FLUX simulations for 2 MeV protons along the Si <111> axis showing the relationship between spatial distribution (upper row) and the corresponding angular distribution (lower row), for path lengths of 95 nm and 130 nm, corresponding respectively to the large peak and subsequent trough in the NEP. The saturation level in the angular images is medium.

Consider the regimes of large $b$, from 0.4 to 0.6 Å and 0.2 to 0.4 Å, incident on a central string. Smaller $b$ gives a larger deflection angle with ions moving further away from a central string. The focusing effect by the first ring depends on $b$; for $b$ = 0.2 to 0.4 Å the higher radial energy means that this component is only weakly focused in passing through the first radial ring, Fig. 5a, resulting in intense, localized angular peaks. It becomes well focused as it interacts with the second ring, Fig. 5b, where it is strongly scattered in the azimuthal direction, producing a ring-like distribution at ~0.6$\psi_a$.

As a general observation, as a beam component passes through a potential minima between strings its radial angle is increased and where it approaches a potential maxima around strings its radial angle is decreased. The resolved angular peaks observed at ~0.6$\psi_a$ can thus be explained as follows. Fig. 6 shows spatial and angular maps for the combined range of $b$ = 0.2 to 0.6 Å at a depth of 95 nm. The smaller $b$ components produce the outer angular peaks where most ions have passed through the first radial ring so their decreased potential energy gives a greater radial angle. The larger $b$ components produce the inner angular region where most ions are close to the first radial ring and their increased potential energy gives smaller radial angle. The resolved dots therefore arise from the potential energy barrier in passing through a radial ring. They are experimentally observed in Fig. 4b since the $b$ components involved are relatively large, with ~21% of the total beam having $b$ < 0.6 Å.

*3.3. Small impact parameters b < 0.2 Å*

The large components of $b$ considered above produce deflection angles up to ~0.6$\psi_a$ and the characteristic angular structure is similar to that observed in experimental patterns within this angular range. For $b$ < 0.2 Å a continuum of deflection angles greater than ~0.6$\psi_a$ is produced and several focusing conditions can be simultaneously observed at different radial rings. The deflection angle increases with smaller $b$ but with decreasing frequency so the deflected intensity reduces with radial distance/angle away from the central string/axis. Here we consider the higher intensity effects produced by the larger values of $b$ in the range of $b$ < 0.2 Å since Fig. 4b demonstrates that it can be clearly observed up to ~0.8$\psi_a$. The lower intensity, larger deflections produced by $b$ < $\rho_{th}$ are considered in section 3.4.

Fig. 5 shows the component $b < 0.2$ Å at the same depths and saturation as the other $b$ components. At a depth of 130 nm there are two ring-like distributions arising from different radial focusing conditions met by different $b$ components within this range. The inner ring at ~$0.6\psi_a$ is the same as that shown for $b = 0.2$ to 0.4 Å which scatters from the second radial ring. The outer ring at ~$0.8\psi_a$ is formed by smaller $b$ components being focused onto, and scattered by the ring 4, as discussed below.

Spatial and angular maps in Fig. S2 show the full range of depths with small increments and at several saturation levels, providing the best visualization of how angular and spatial distributions evolve for $b < 0.2$ Å. Several interesting effects are seen which are not discussed here. Fig. 7 shows a selection of these spatial and angular maps for the same depths as Fig. 3 to observe how this $b$ component propagates through successive radial rings. The following discussion refers to the motion of the more intense beam seen in the lower saturation images. Beam deflected away from the central string interacts with the first radial ring (depth of 40 nm) giving the first, small NEP peak. The higher radial energy of this $b$ component, compared to those for $b > 0.2$ Å, results in it being steered between strings of the second and third radial rings (depths of 60, 80 nm), producing a focusing effect on the angular distribution. This component is focused onto strings in the fourth radial ring (depth of 95 nm), giving the second, large peak in the NEP. The resulting strong scattering deflects the focused beam to either side of the strings, resulting in a large increase in azimuthal angle so the angular peaks are split; by a depth of 130 nm at the trough after the second NEP peak the angular distribution has evolved into another ring-like shape at ~$0.8\psi_a$.

A general observation regarding the relationship between the NEP depth distribution and the spatial and angular distribution is that where beam is steered or focused between strings (i.e. NEP is a minimum) then angular peaks are produced. Where well-focused beam components are strongly scattered by a radial ring (i.e. NEP is a maximum) the angular peaks are split into two and then the azimuthal angular range is greatly increased, forming a ring-like distribution. Thus, NEP peaks and troughs can be inferred from observation of characteristic angular structure in channeling patterns in the range of 0.6 to $0.8\psi_a$ since this range is produced by $b < 0.2$ Å which was shown to dominate the NEP in Fig. 1.

Consider motion at depths beyond the NEP trough at 130 nm. After being scattered from the fourth radial ring, about half the ions are deflected towards ring 6 where strings are aligned with {112} directions. The resultant scattering produces the third NEP peak at a depth of 160 nm. Ions are not well focused onto ring 6, making only glancing angle scattering so they to continue their motion close to {112} directions beyond this depth. The other half of the ions are scattered from ring 4 towards {110} directions where they also make a series a glancing angle collisions and continue their motion close to this direction.

The process whereby ions become aligned with major planar directions continues with depth so the spatial distribution becomes strongly peaked along these directions, see Fig. 8a at depth of 350 nm. Fig. 8b shows a simulated trajectory plot through the same depth where the location of strings and the first eleven radial rings are indicated. It is now clear how the overall effect of radial focusing, steering and scattering of small $b$ components results in preferential alignment towards major planar directions where they undergo a successive scattering from strings along these directions.

The different contributions of each ring on the component $b < 0.2$ Å can be summarized as follows. The first two rings, by definition, comprise strings running along major planar directions which intersect the aligned axis, i.e. {110} and {112} in this case. Their main effect is to steer and focus beam onto rings 4 (which produces the second, large NEP peak) and then rings 6 and 7 which produce the third NEP peak. Beyond this depth many ions are aligned towards major planar directions.

*3.3.1. Phase Space Analysis*
Fig. 8d shows a phase space plot at a depth of 350 nm for the angular region extracted from within the dashed box in the corresponding angular image in Fig. 8c, starting from 0.7° away from the <111> axis, i.e. significantly beyond $\psi_a$. See Refs. [27,28] for discussions of phase space analysis of planar channeling. On the vertical axis the phase space plot shows the angular coordinate perpendicular to the vertically-running {110} direction ($\psi_p = 0.17°$) versus the position coordinate perpendicular to the vertically-running {110} direction on the horizontal axis. Ten adjacent {110} planes are shown, equally spaced about the central string, location indicated by a white arrow. A well-channeled planar beam would occupy phase space coordinates within the black ellipses indicated, such that where they are located at the mid-point between channel walls their transverse angle is $< \psi_p$ and where they are located close to the channel walls their transverse angle approaches zero. Conversely beam which is blocked from a planar direction occupies phase space coordinates outside the ellipses, so that ions which are located at the mid-point between channel walls have a transverse angle $> \psi_p$ and those which are located close to the channel walls have a transverse angle which is sufficiently large so that they overcome the planar potential barrier. The phase space plot shows that along the {110} planes containing the central string on which all ions are incident, a significant number of trajectories are within the ellipse though lying close to its boundary, indicating large amplitude planar trajectories. While these are rapidly dechanneled other trajectories may be scattered into a planar channeled trajectory. Note the similarity between the NEP oscillations in Fig. 1 for $b < 0.2$ Å and those for a {110} planar channeled beam; they have a similar oscillation period and a similar decreasing amplitude of oscillation. We conclude that there is a significant component of protons scattered with small impact parameter which undergoes large amplitude planar oscillations even at axial alignment. They produce small oscillations in the NEP which are obscured by larger $b$ components.

The above analysis provides the mechanism underlying "*scattering in the transverse plane*" as elucidated by Armstrong et al [19] as follows: as channeled ions travels through the crystal they collide with many atomic strings, acquiring different transverse momentum with successive collisions. The azimuthal structure depends on the extent to which the transverse proton momentum is rotated from the initial direction; some azimuthal directions are favoured, producing intensity variations around the ring. Less intense azimuthal directions correspond to intersecting major planar directions, i.e. when ions are scattered so that their momentum is directed these directions in the transverse plane, they will be blocked and the transverse momentum rotated away from these directions.

In Fig. 8 we see that this process involves a correlated sequence of steering and focusing by successive radial rings which orient trajectories towards major planar directions, producing large amplitude planar channeling trajectories which travel much further radially than other azimuthal directions.

*3.3.2 Multiple Reflections*

Now consider those ions which undergo large angle deflections by radial rings so they do not continue moving away from the central string. They may be deflected sideways or reflected back towards the central string, after which they may again be focused in any direction. This may occur several times, producing a multiplicity of beam propagation directions within the simulated patterns. Considering the more-saturated simulated spatial images in Fig. 7, for a depth of 40 nm the region around the central string is empty as all trajectories move outwards. However, by a depth of 80 nm some ions are reflected from the first radial ring and may propagate back towards the central string where they are again reflected and move outwards again. For all greater depths that there are low intensity regions within the inner radial rings which contain an intricate network of multiple reflections.

Barrett implicitly assumed that in characterizing the effect of interactions with radial rings that most ions continue in an outwards direction, though it is noted that in his simulations for the lowest radial energy component, Fig. 3 in Ref. [10], that most trajectories do undergo large angle deflections. Such multiple reflections are the likely explanation as to why his simulations showed NEP contributions from the inner rings at greater depths, through which most ions with high radial energy had already passed.

*3.4 Impact parameters $b < \rho_{th}$*
In Fig. 5a the simulated angular distribution for $b < (\rho_{th} = 0.106$ Å$)$ exhibits angular structure within a range of deflections from ~0.9 to a maximum of ~1.3$\psi_a$, as indicated by an arrow. This range of deflection angles was only faintly visible in simulations for the full beam distribution in Fig. 3c, owing to the small beam fraction contributing to it. Further reducing $b$ in simulations increases the frequency of large deflections so in Fig. 5a for $b < 0.01$ Å angular structure up to ~1.6$\psi_a$ is seen.

One can observe the same faint angular structure from ~0.9 to 1.3$\psi_a$ in the more saturated images Fig. 7 where the angular and spatial evolution of ions undergoing such small $b$ collisions can be followed with depth. In Fig. 7a at depths of 60 to 80 nm this component is poorly focused onto ring 4, resulting in scattering over a narrow azimuthal angular range and radiating angular lines. These are similar to those for the slightly larger $b$ components at the same depth in the lower saturation image which were produced from poorly focused beam deflected past ring 2. This component is steered towards ring 7 (rather than ring 6 for slightly larger $b$) and scattering from this ring at depths of 90 to 100 nm produces the outer arrowed azimuthal circular distribution. Therefore two radial focusing conditions, from rings 4 and 7, are simultaneously met by different components of $b < 0.2$ Å. At greater depths the outcome is the same as for larger $b$; ions are scattered towards major planar directions with subsequent interactions being dominated by successive collisions with strings along them. Concentric angular structure beyond ~0.9$\psi_a$ is therefore lost beyond depths of 100 nm.

*3.5. Small tilts away from <111> axis*
Fig. 9a shows experimental channeling patterns of 2 MeV protons transmitted through a 95 nm Si layer for different tilts to the <111> axis along the vertically-running (011) direction. Fig. 9b shows two corresponding simulated angular distributions. Tilting imparts radial energy to the full beam distribution so an intense azimuthal ring is formed, concentric with the beam axis and a radius equal to the tilt. This "doughnut" distribution was noted by others in thicker Si membranes [19,20]. In our ultra-thin membranes the azimuthal ring is surrounded by a second, faint, outer region comprising the same angular structure as observed at axial alignment (see inset for a tilt of 0.15°). The angular structure within the outer region does not significantly change with tilt but its intensity does, with the lower/upper halves becoming more/less intense.

At a tilt of 0.25° an azimuthal ring of peaks is observed oriented along {112} directions, similar to those observed at axial alignment (see inset). Since the main beam distribution is now located on the azimuthal ring, such angular features are more easily resolved than at axial alignment. They arise from a similar focusing effect by atomic strings, Fig. 9c. When considering the evolution of the full beam distribution one cannot consider scattering only from a central string, however it suffices to consider radial motion outwards from the incident unit cell. Many ions are focused towards {112} directions, indicated by white arrows in Fig. 9c. Consider motion shown within the white ellipse along the horizontally-running {112} direction. Many ions are at potential minima between strings giving isolated angular dots, similar to those in Fig. 7a, depth of 60 nm. For higher radial energy (tilt of 0.30°) ions are steered either side of the row of strings within the ellipse, producing radial angular lines similar to those in Fig. 7a, depth of 80 nm.

Now consider the effect of tilting in revealing faint characteristic angular structure at large deflection angles, up to ~1.3$\psi_a$. In Fig. 9 for tilts of 0.25° and 0.30° a white arrow indicates the same characteristic triangular feature in this angular range as in Fig 5a. This feature is now observable experimentally owing to the tilting effect which provides a larger beam fraction with large angular deflections. This confirms the predicted existence of such angular structure beyond $\psi_a$. Two further simulation sequences are included in Figs. S3 and S4, and experiment sequences in Fig. S5 showing the evolution of the doughnut distribution with different (011) tilts and path lengths.

### 3.6. Large tilts beyond $\psi_a$ from <111> axis

Since the early work of Bøgh (chapter 15 of Ref. [2]), other groups have reported anomalous peaks in RBS-channeling spectra at certain planar and axial alignments using high-resolution magnetic spectrometer and standard surface barrier detectors respectively where the transverse focused beam strongly interacts with neighbouring atomic strings or planes. Dygo measured RBS spectra for MeV helium ions incident on GaP [29] and Si [30] versus tilt away from the <110> axis along the (110) direction and observed several peaks. Smulders et al [31] measured RBS spectra versus tilt away from the <110> axis along the (211) direction where a large peak height of 2.5 times the random level was observed at tilts.

Fig. 1 shows the NEP at <111> axial alignment with peaks at depths corresponding to where the beam components of $b < 0.2$ Å are focused and scattered from radial rings, establishing a connection between the NEP and the angular distribution at that depth. In Fig. 10a the NEP depth distribution along the {011} direction remains almost the same for tilts up to 0.2°, though increasing in amplitude. At larger tilts strong NEP oscillations are observed. We attribute this to the radial focusing effect whereby beam is focused and steered towards the vertically-running {011} directions, in a similar manner as shown in Fig. 8b, resulting in pronounced planar oscillations. Tilting increases the beam fraction involved beyond that for the small $b$ component at axial alignment, making the effect more obvious.

This insight as to how the radial rings focus and steer beam towards those strings which are aligned with minor planar directions, resulting in strong scattering, can be used to predict general tilting behavior along all such minor planar directions. Fig. 10b shows the NEP for tilts along the {123} direction, corresponding to ring 4 which is responsible for the large NEP peak at axial alignment. Tilting causes a large beam fraction to become focused on to ring 4, resulting in a very large peak in the NEP for a tilt of 0.55° (~1.3$\psi_a$).

The upper plots in Fig. 10 show maps of the evolution of the NEP depth distribution with tilt angle. For the {011} direction, vertical (light blue) bands correspond to the planar oscillations which are present for all tilts beyond of 0.2°. Similar, fainter vertical bands are also seen for the {123} direction, arising due to scattering from ring 4 directing a large beam fraction towards the {011} direction where it undergoes planar oscillations. We consider that the fine structure observed in the simulated NEP maps of depth versus tilt angle in Fig. 10 is produced by multiple beam reflections discussed in section 3.3.2.

Other minor planar directions associated with rings 7, 9, 10 exhibit similar peaks in the NEP on tilting beyond $\psi_a$. Fig. 11 plots the integrated NEP over a depth of 50 nm to 150 nm versus tilt angle away from the <111> axis for all minor and major planar directions indicated in Fig. 2. For shallow depth ranges (tens of nanometers) from the surface there is little variation with tilt since the radial distance travelled is not sufficient to interact with the radial ring along that direction. A depth range up to 150 nm results in strong variations with tilt angle since ions have travelled through a radial distance sufficient to interact with the relevant ring. Note the large difference in tilt at which maxima of shoulders are located, ranging from ~0.55° along the {123} direction (ring 4), to ~1.1° along the {156} direction (ring 10). This is consistent with the difference in radial distance travelled to interact with these radial rings, equal to ~4.5 Å at a depth of 50 nm, comparable to the difference in radii of the fourth and tenth radial rings. Beyond 150 nm the angular variation reduces since little of interest occurs in the NEP depth distribution beyond this depth.

This effect underlies previous observations of peaks in RBS channeling spectra on tilting away from axial alignment, being manifested most strongly for tilts along minor directions corresponding to the same inner rings on to which beam is focused at axial alignment, and over a narrow depth interval beneath the surface. Furthermore it explains how large shoulders are formed on RBS channeling angular scans at tilts beyond $\psi_a$ and why the height of these shoulders is different if one tilts along a different radial direction. This effect is analogous to that described by Wijesundera et al [32,33] where large shoulders observed on tilts perpendicular to major planar directions are shown to be due to beam being spatially focused into the walls of the channels at certain depths.

### 3.7. Channeling patterns at the <001> axis
So far we have only discussed effects along the <111> axis where radial motion due to small $b$ scattering is clearly observed. Similar effects are observed at other major axes. Fig. 12a shows a potential map of the Si <001> axis. For 1 MeV protons in Si [001] $\psi_a = 0.50°$ and $\psi_{1/2} = 0.42°$.

Fig. 12b shows the NEP depth distribution for 1 MeV protons aligned with the <001> axis. Similar oscillations are observed to that along the <111> axis and a similar relationship between NEP peaks/troughs and angular structure is observed in channeling patterns. Fig. 12c,d show simulated (full beam distribution) and experimental patterns for 1 MeV protons aligned with the surface-normal [001] axis of a 55 nm thick membrane, i.e. just before the large NEP peak. The central cross produced by large impact parameter ions has been the subject of a number of studies on Rainbow channeling and superfocusing effects [17]. The faint lines extending away from the central cross, with more intense, resolved angular peaks at their outer extent are produced by a similar mechanism as described in Fig. 6, whereby ions moving beyond the first radial ring are influenced by the potential around the second ring, introducing a gap in the deflection angle. Fig. 12e,f show results for a 100 nm thick membrane, i.e. a path length corresponding to the trough after the large NEP peak. Scattering from the second ring results in the beam becoming spread out azimuthally, but now it forms a square shape, rather than circular shape as at the <111> axis, owing to the different symmetry of the <001> axis.

Fig. S6 shows how the beam component for $b < 0.2$ Å propagates radially outwards versus <001> layer depth and the spatial and angular maps exhibit similar features as the <111> axis and can be explained in the same manner.

### 4. Discussion & Conclusions

We have shown how various phenomena associated with ion channeling in thin and ultra-thin crystals can be understood within the framework of Barrett's radial focusing effect by concentric rings of atomic strings. This includes showing that different impact parameters can be resolved by choosing specific angular regions of the exit angular distribution and related to the NEP depth distribution, and how angular peaks are produced around azimuthal ring of the doughnut distribution for small tilts away from axis. It was also shown how planar channeling of small $b$ components can arise at axial alignment; we note that in Ref. [34] analytical calculations were used to predict a similar effect whereby for ion trajectories with radial high energies a transition from axial channeling to planar channeling was possible.

This study is based on Barrett's earlier work [10] on radial focusing in <111> tungsten and it is worth considering why some of our observations were not found in his original work. Barrett studied radial focusing over a range of radial energies below and above the maximum potential energy of the atom strings and summed them together to find the overall contribution to the NEP. He calculated the NEP from each of the first eleven radial rings and combined together the contribution from outer rings. He found that beam was focused on to rings 4 and 7 and the resulting strong scattering produced peaks in the NEP. Contributions to the NEP due to different $b$ components were not considered so oscillations in the NEP arising from small $b$ components at depths beyond the first three peaks were not found. Since only spatial focusing was studied, the angular distribution was not considered so a link between characteristic features observable in channeling patterns at peaks and troughs in the NEP was not possible. It is also worth considering why such effects were not manifested in Barrett's earlier simulation study on location of shoulders in channeling phenomena, where he studied the nuclear reaction probability of 1.4 MeV protons in aluminium on tilting away from the [011] axis, see Figs. 2, 3 of Ref. [8]. The most likely explanation is that he used a layer thickness of 29 nm, which is too thin to include the large NEP depth peaks, as Figs. 10 and 11 demonstrate.

This study is also relevant to work on deflecting very high energy proton beams (hundreds of GeV) in bent Si crystals. Beam aligned with the <111> entrance surface [35] was deflected and split into two "skew" channels. The detailed mechanism for this is likely to be similar to that presented here whereby a beam fraction is scattered into planar channels in straight crystals by examining motion in successive radial rings; a bent crystal may produce similar, stronger effects along certain planar directions.

**Figure Captions**

**Figure 1.** NEP versus depth for 2 MeV protons along the Si <111> axis, for the full beam distribution and components of $b$, accounting for the small beam fraction incident within these ranges. 5 million trajectories were simulated in each case. Also shown for comparison is the NEP produced by tilting away from the <111> axes along the (110) direction by 2°, where planar oscillations result in large oscillations. The vertical scale is compressed for comparison with the others.

**Figure 2.** Static atomic potential averaged along the ion direction at the Si <111> axis. Red regions represent low potential in the scale bar. Numbered radial rings of atomic strings away from the central string, located in the bottom left corner are shown. Rings 1, 3, 5, 8, 11 (orange colour) have atom strings aligned along {110} directions, rings 2,6,12 (black colour) are aligned with {112} directions. Purple coloured rings correspond to those aligned with minor planar directions.

**Figure 3.** Simulated channeling patterns of 2 MeV protons transmitted along the Si <111> axis for different path lengths. Three image saturations are (a) medium, (b) high, (c) very high. Where relevant the extent of $\psi_a$ and $0.5\psi_a$ are shown by dashed circles.

**Figure 4.** Experimental channeling patterns for (a) 1 MeV, (b) 2 MeV protons along the Si <111> axis for a path length of 95 nm. The extent of $\psi_a$ is shown by a dashed circle.

**Figure 5.** Simulated angular and spatial distributions for 2 MeV protons transmitted along the Si <111> axis for a path length of (a) 95 nm and (b) 130 nm, for different ranges of $b$. The angular extent of $\psi_a$ is indicated by a dashed circle and atomic strings are indicated as white dots.

**Figure 6.** Simulated spatial and angular distributions for 2 MeV protons along the Si <111> axis for a path length of 95 nm for $b < 0.2$ to 0.6 Å. The angular extent of $\psi_a$ is indicated by a dashed circle and atomic strings are indicated as white dots.

**Figure 7.** Simulated angular and spatial distributions for 2 MeV protons transmitted through an aligned Si <111> membrane for different path lengths for $b < 0.2$ Å. Two image saturations (medium and high) are shown so as to highlight different beam components. The angular extent of $\psi_a$ is indicated by a dashed circle and atomic strings are indicated as white dots.

**Figure 8.** Simulated (a) spatial and (c) angular distributions for 2 MeV protons transmitted through an aligned Si <111> membrane for a path length of 350 nm. (b) Corresponding trajectory plot at a depth of 350 nm depth over an area of 10 Å × 10 Å as indicated in (a), with the central string in the lower left corner and trajectories in the upper right quadrant plotted. (d) Corresponding phase space plot at 350 nm depth, based on trajectories within the dashed box in (c).

**Figure 9.** (a) Channeling patterns recorded for 2 MeV protons transmitted through a 95 nm Si <111> membrane for different tilts along the vertically-running {011} direction. (b), (c) Corresponding FLUX simulations of the angular and spatial distribution respectively for the full beam distribution. The inset for tilts of 0.15° and 0.25° respectively show the high and medium saturation images from Fig. 4b.

**Figure 10.** NEP depth distribution for tilts away from the <111> axis along the (a) {011}, (b) {123} directions. The lower plots show the NEP

to a depth of 500 nm at specific tilts, where the absolute maxima can be compared (note the difference in vertical scales) between (a) and (b). The upper maps plot the NEP to a depth of 500 nm on the horizontal axis and tilt angle on the vertical axis.

**Figure 11.** Integrated NEP over the depth window from 50 to 150 nm for radial tilts away from the <111> axis along major and minor planar directions.

**Figure 12.** (a) Potential map of Si <001> axis and (b) NEP depth distribution for 1 MeV protons. (c),(e) simulated angular distributions for the full beam distribution for 1 MeV protons transmitted through an aligned 55 nm and 100 nm thick Si <001> membrane. (d),(f) Corresponding experimental channeling patterns. The angular extent of $\psi_a$ is indicated by a dashed circle where necessary.

**Figure 1.**

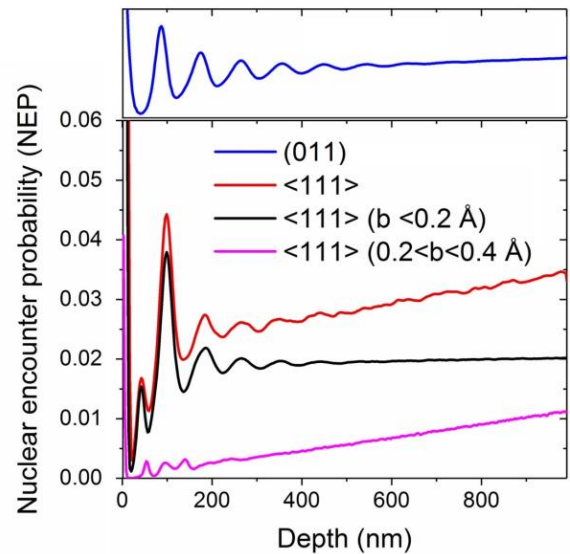

Figure 2.

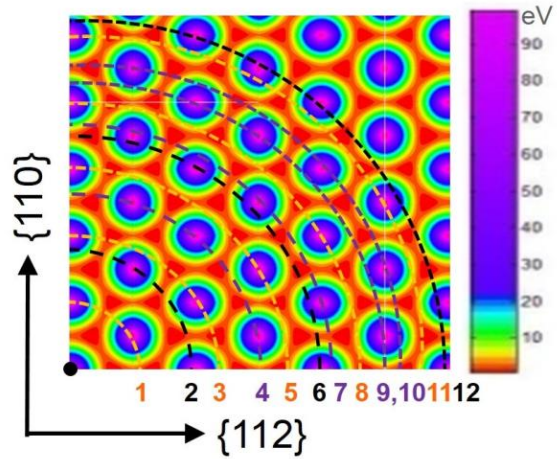

Figure 3.

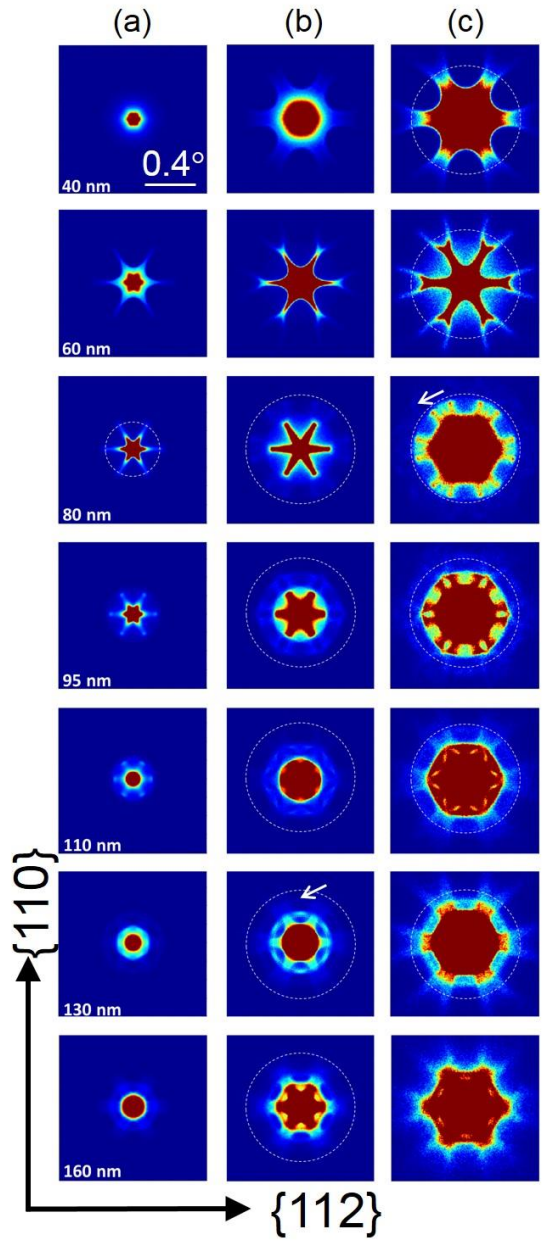

Figure 4.

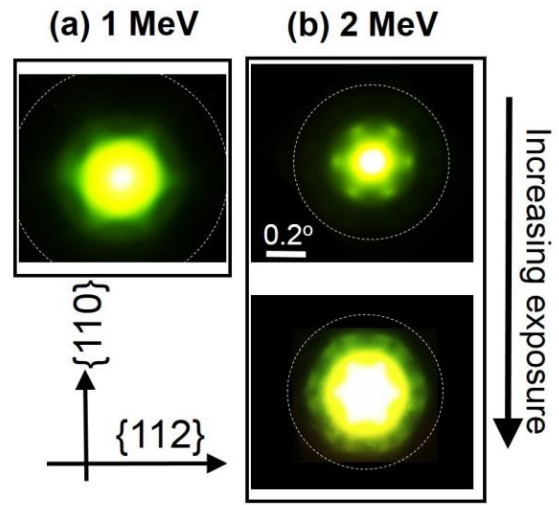

Figure 5.

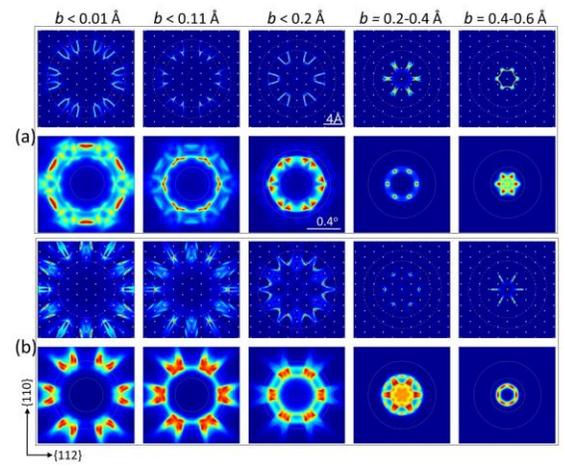

Figure 6.

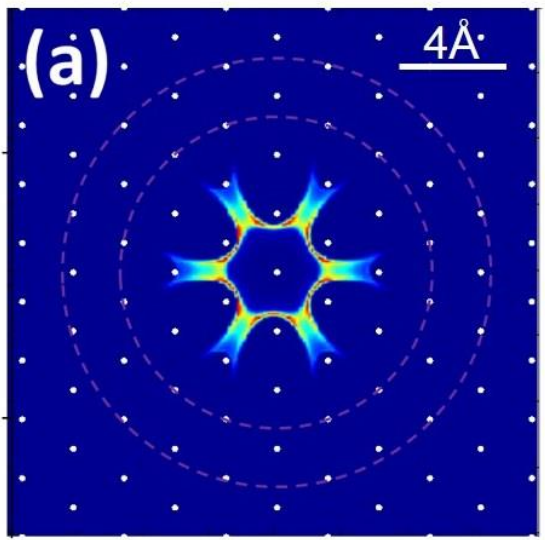
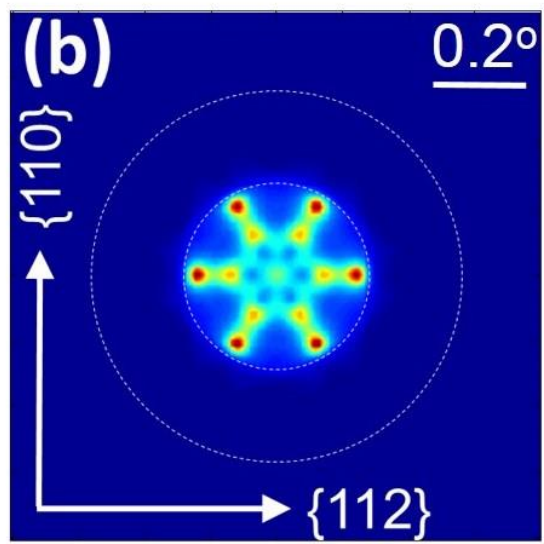

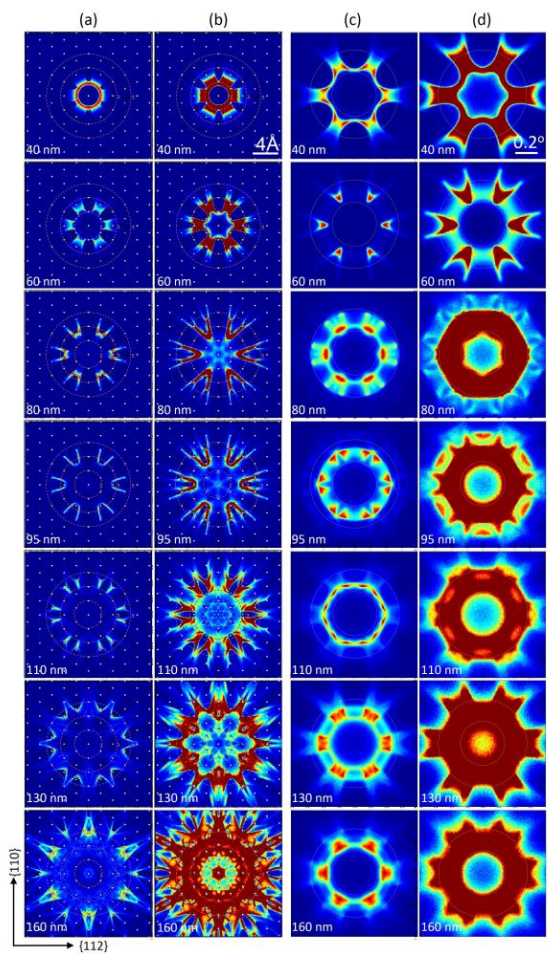

Figure 7.

Figure 8.

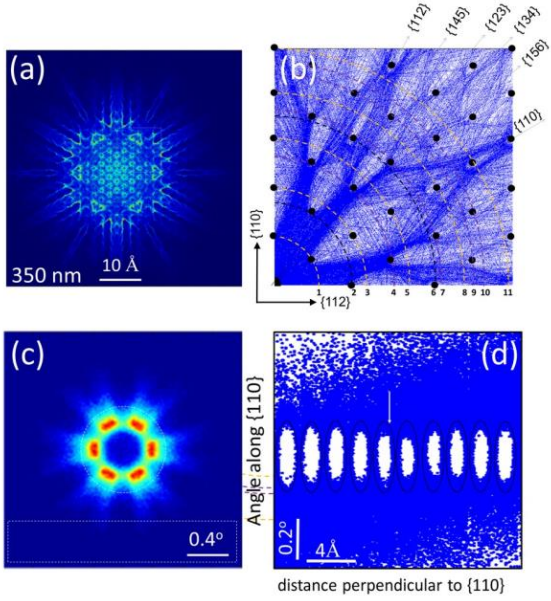

Figure 9.

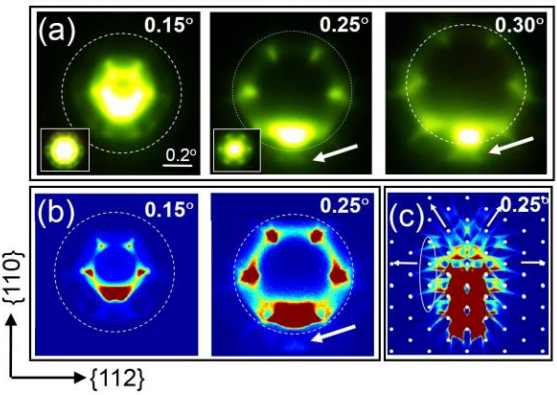

Figure 10.

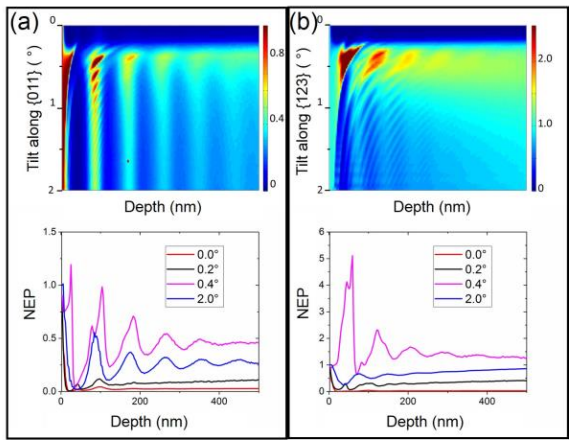

Figure 11.

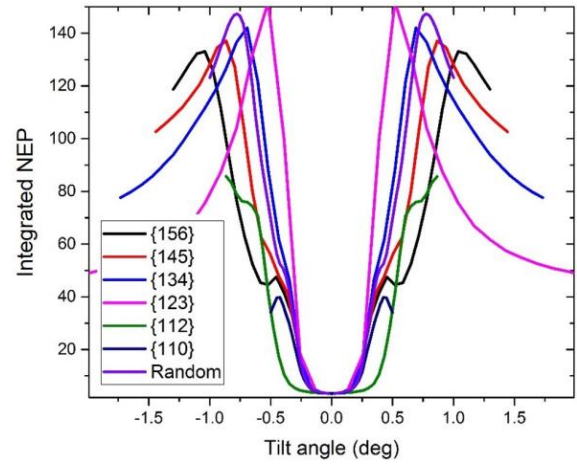

Figure 12.

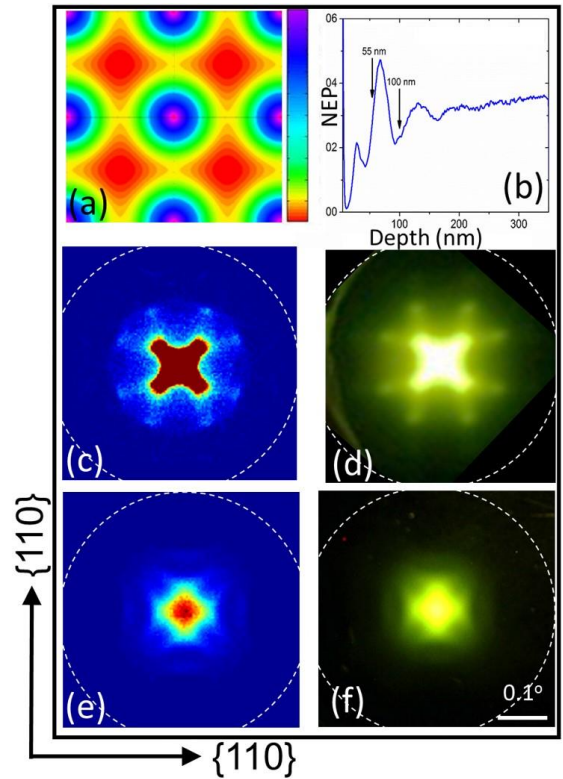